# Valley and pseudospin-valley topologically protected edge states in symmetric pillared phononic crystals

**Wei Wang[1], Bernard Bonello[1*], Bahram Djafari-Rouhani[2], and Yan Pennec[2]**

[1]Sorbonne Université, UPMC Université Paris 06 (INSP–UMR CNRS 7588),

4, place Jussieu 75005 Paris, France

[2]Institut d'Electronique, de Micro-électronique et de Nanotechnologie (IEMN–UMR CNRS 8520),

Université de Lille Sciences et Technologies, Cité Scientifique, 59652 Villeneuve d'Ascq Cedex, France

*corresponding author: bernard.bonello@insp.jussieu.fr

Abstract:

We present a symmetric double-sided pillared phononic crystals (PPnCs) that can emulate both quantum spin Hall effect (QSHE) and quantum valley Hall effect (QVHE) by solely imposing different geometric perturbations. Indeed, the Dirac cones can occur in the low (deep subwavelength) and high frequency regime by judiciously turning the parameters of the attached pillars and even a double Dirac cone can be achieved. We realize the valley-protected, the pseudospin-protected or the pseudospin-valley coupled edge states with the proposed platform. Besides, we show a variety of refraction phenomena (positive, negative and evanescent) of the valley-polarized edge state at the zigzag termination when emulating QVHE. Further, we illustrate the valley-dependent feature of the pseudospin-valley coupled edge state and demonstrate the valley based splitting of the pseudospin-protected edge states in a Y-junction wave guide.

The desire to manipulate elastic wave, especially Lamb waves, in phononic crystals (PnCs) by tailoring the band structures from either band folding or local resonances has been one of the most interesting pursuits in the past decades. Abnormal wave propagation phenomena, such as negative refraction [1,2], lensing [3,4] and clocking [5,6] unachievable in natural materials has been demonstrated. Recently, the emergence of topological insulators labelled as a new state of matter in condensed-matter physics provides a fascinating approach to achieve the defect-immune and lossless energy transport [7–10]. The concept has been quickly extended into PnCs to realize the robust guiding of Lamb waves in analogy to quantum spin Hall effect (QSHE) [11–18] or quantum valley Hall effect (QVHE) [19–23]. In the former case, it explores the pseudospin degree of freedom and needs intricate designs to obtain a double Dirac cone at Γ or K (K') points of the Brillouin zone (BZ). Usually, for the one occurring at Γ point, it can be realized by employing the zone folding mechanism [24,25], whereas for the one occurring at K (K') points, specifically patterned plate with optimized parameters are required [26,27]. While the latter case explores the valley degree of freedom and only needs single Dirac cone at K (K') points of BZ which can simplify the design and fabrication significantly, nonetheless, the large separation between two opposite valleys should be ensured to avoid the inter-valley scattering. Up to now, the Dirac cone occurring at the deep subwavelength scale has seldom been investigated and few attention has been paid to the refracted pattern of the valley-polarized edge state at the zigzag terminations.

Very recently, the topological edge state that combines both the pseudospin and valley degrees of freedom has gained more interests. It has been reported that the pseudospin-valley coupled edge states can occur at the domain wall formed by two distinct configurations supporting QSHE and QVHE respectively [28–34]. However, these two configurations are usually completely different in geometry that adds complexity. Therefore, to figure out a platform that can transit between both QSHE and QVHE plays an important role in this research. In this Letter, we present a symmetric double-sided pillared phononic crystal (PPnC) that can emulate both QVHE and QSHE by solely imposing different geometric perturbations. By appropriately choosing the parameters of the attached pillars, the Dirac cones can occur in the low (deep subwavelength) and high frequency regime and even a double Dirac cone can be obtained. We demonstrate the occurrence of the valley-protected, the pseudospin-protected or the pseudospin-valley coupled edge states at different domain walls. When emulating QVHE, we show a variety of refraction phenomena of the valley-polarized edge state at the zigzag termination, including positive, negative and evanescent. Further, we illustrate the valley-dependent feature of the pseudospin-valley coupled edge state and demonstrate the valley based splitting of the pseudospin-protected edge states in a Y-junction wave guide.

The elementary unit cell of the proposed PPnC constructed by assembling two identical periodic arrays of pillars over a thin plate and its first irreducible BZ are shown in Fig. 1(a). Four perforated holes are drilled at the corners of the honeycomb unit cell. The lattice constant and the thickness of the plate are chosen to be $a = 400\mu m$ and $e = 100\mu m$. The diameter and height of the pillar are $d_A = d_B = d_C = d_D = D = 120\mu m$ and $h_A = h_B = h_C = h_D = H = 160\mu m$. Regarding the perforated holes, the diameters are set to be $d_H = 280\mu m$. In the following discussion, both the plate and the pillars are chosen to be steel whose Young's modulus, Poisson's ratio and mass density are $E = 200GPa$, $v = 0.3$ and $\rho = 7850kg/m^3$. Note that the reported phenomena apply to any other single-phased PPnCs.

By applying periodic conditions and solving the eigenvalue equations, the band structure of the double-sided PPnC is obtained and displayed in Fig. 1(b). Owing to the symmetry about the mid-plane of the plate, it can be further decomposed into the symmetric (red dotted lines) and antisymmetric (blue dotted lines) subcomponents (see the eigenmodes in Fig. S1 [35]). For comparison, the band structure of the single-sided PPnC, by removing the pillars on one side, is shown in Fig. 1(c). Two Dirac cones formed by the antisymmetric modes persist. Whereas the one created by the symmetric modes disappears and at the degenerate frequency the wavelengths in the plate are 2557μm for $S_0$ mode and 1513μm for $SH_0$ mode. Obviously, it occurs at the deep subwavelength scale and can be attributed to the local resonances of the attached pillars. Note that both Dirac cones can be judiciously tuned (Fig. S2 [35]). Here, we concentrate on the one of the symmetric dispersion curves shown in Fig. 1(b) and the others are analyzed in [35].

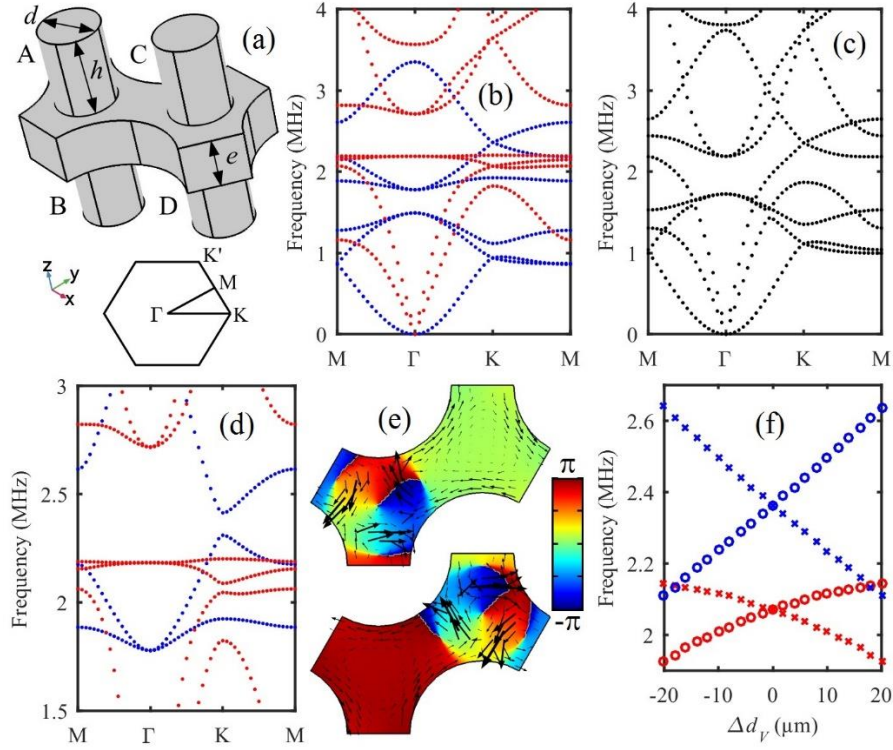

FIG. 1: (a) Elemental unit cell and its first irreducible BZ. Band structures of (b) the double-sided, (c) the single-sided and (d) the perturbed PPnCs with $\Delta d_V$ = 4μm. (e) Energy flux (black arrows) and phase distribution at the lower (bottom panel) and higher (top panel) bounding of the lifted Dirac cone at the valley K. (f) Evolution of the valley pseudospin states at the valley K against the perturbation. Red (blue) symbols represent the states of the symmetric (antisymmetric) dispersion curves.

To emulate QVHE, the perturbation in the diameter of the pillar with $d_A = d_B = D + \Delta d_V$ and $d_C = d_D = D - \Delta d_V$ is introduced to break the space-inversion symmetry (SIS) in the unit cell. Specifically, when $\Delta d_V$ = 4μm that belongs to the small SIS breaking case, the band structure is displayed in Fig. 1(d). Both Dirac cones are lifted that leads to two nontrivial band gaps. For the symmetric (antisymmetric) dispersion curves, both the energy flux and the phase distribution on the top surface of the plate at the lifted Dirac cone are depicted in Fig. 1(e) (Fig. S3(a) [35]) which unambiguously reveals the vortex chirality and can be recognized as the valley pseudospin states. Figure. 1(f) shows the evolution of the valley pseudospin states against the perturbation. The cross (circle) symbols represent the valley pseudospin down (up) states. When crossing zero, their frequency order is inverted and the band gap will firstly close and then reopen again that indicates the topological phase transition. Then, the valley Chern number becomes nonzero. The numerical integration of the Berry curvature in the vicinity of the valley K is in agreement with the theoretical value ±1/2 (see Figs. S3(b) and S3(c) [35] for detailed results). A discrepancy usually occurs in the strong SIS breaking case [36]. Upon the same perturbation, the band gap between the valley pseudospin states of the antisymmetric dispersion curves is much broader which suggests that it is more sensitive to the perturbation in the diameter. However, it is less sensitive to the perturbation in the height (Fig. S3(d) [35]).

To investigate the edge states, a three-layer ribbon supercell illustrated in Fig. 2(a) is considered. Two different zigzag domain walls are established by arranging pillars with decreased (SDW) and increased (LDW) diameters adjacently. Three branches appear in the reopened band gap as shown in Fig. 2(b) and the eigenmodes at $k_x = 0.5\pi/a$ are displayed in Fig. 2(b) in the descending order. The magenta and cyan dotted lines represent the edge states occurring at SDW and LDW respectively and the black dotted line depicts the locally resonant mode at the bottom end where the pillar with increased diameter is placed outside (see the zoomed view in the inset). Subsequently, the propagation of the K'-polarized edge state at SDW and its refraction at the zigzag termination is investigated. The left-going wave at 2.042MHz is launched by two phase-matched sources [37] in the middle. Figure. 2(c) plots the magnitude of the in-plane displacement on the top surface of the plate (see Fig. S7 [35] for the plot of the out-of-plane displacement). The edge state turns downwards and is localized at the interface. It is mainly SH-polarized according to the polarization vectors (black arrows). Part of the energy diffuses into the plate and a small portion propagates along the bottom interface. To interpret this, the equifrequency contours (EFCs) analysis is carried out as depicted in the inset. The red and green solid lines denote EFCs for $S_0$ and $SH_0$ Lamb modes in the plate. Two cyan dashed lines represent the normal to the zigzag termination. The edge state is locked to three K' valleys (black dots) of BZ (black solid line). The wave vector in the plate can be graphically obtained owing to the conservation of the component of the wave vector parallel to the interface [31]. Obviously, it becomes imaginary which means that the refracted wave would be evanescent. However, due to its coupling with the locally resonant mode occurring at the interfaces where the pillars with increased diameter are placed outside (see the zoomed view), the magnitude of the evanescent wave is strongly enhanced. In contrast, the enhanced phenomenon vanishes once the resonant pillars (the highlighted pillars in the zoomed view) are removed as shown in Fig. 2(d) and the magnitude at the bottom interface tends to be zero.

Note that another Dirac cone of the symmetric dispersion curves occurs at the high frequency when decreasing the height of the pillar. Then, the edge state refracts into both $S_0$ and $SH_0$ Lamb waves (see Figs. S10 and S11 [35] for detailed results). More interestingly, for the edge state of the antisymmetric dispersion curves, the refracted pattern consists of one or two $A_0$ Lamb waves at the zigzag termination, whereas one or three $A_0$ Lamb waves at the armchair termination depending on the height of the pillar (see Figs. S4, S5 and S6 [35] for detailed results).

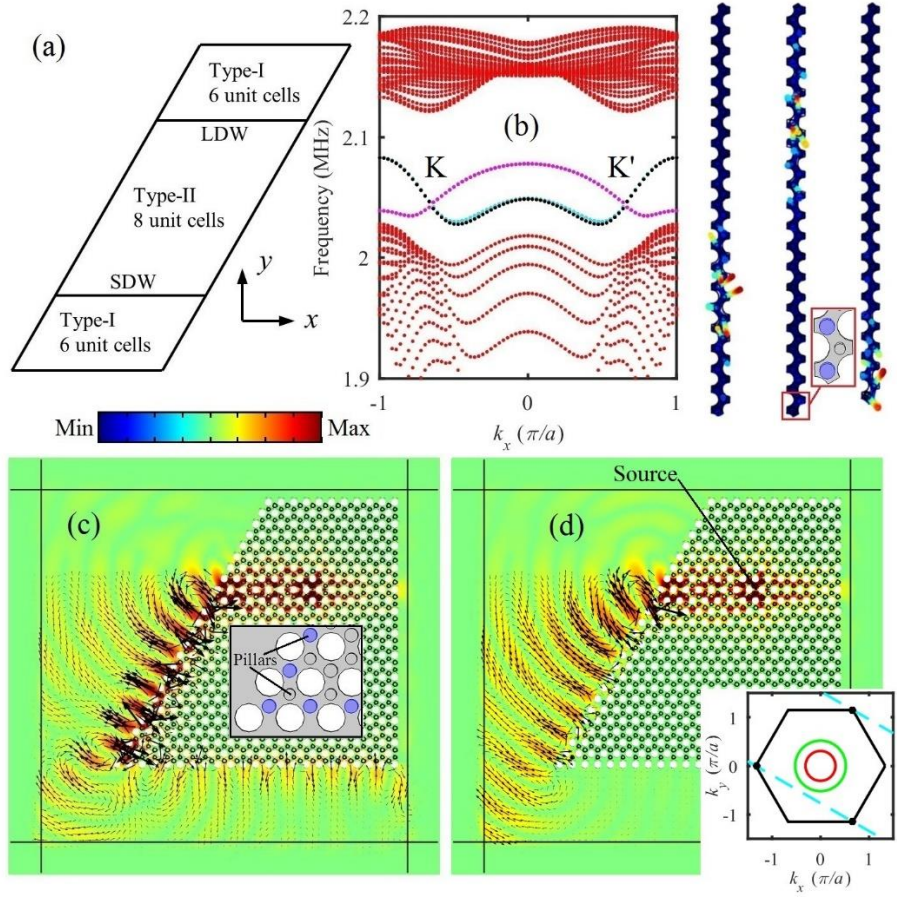

FIG. 2: (a) Schematic of the three-layer ribbon supercell constructed by type-I ($\Delta d_V = 10\mu$m) and type-II ($\Delta d_V = -10\mu$m) PPnCs. (b) Symmetric dispsersion curves of the supercell and the eigenmodes of the magenta, black and cyan branches at $k_x = 0.5\pi/a$ in the descending order. Inset: zoomed view of the undeformed supercell at the bottom end. Plots of the magnitude of the in-plane displacement on the top surface of the plate (c) with and (d) without the resonant pillars at the interfaces. Inset: zoomed view at the left bottom corner and EFCs anlysis.

Recalling the band structures displayed in Figs. 1(a) and S2 [35], the Dirac cones of the symmetric and antisymmetric dispersion curves are close to each other and can be appropriately adjusted. We found that these two Dirac cones can overlay when the parameters are $d_A = d_B = d_C = d_D = D = 80\mu$m, $h_A = h_B = h_C = h_D = H = 153.6\mu$m and $e = 70\mu$m that constructs a double Dirac cone as depicted in Fig. 3(a). Then, it raises the possibility to mimic QSHE. When imposing the perturbation with $h_A = h_D = H + \Delta h_S$ and $h_B = h_C = H - \Delta h_S$, the mirror-symmetry about the mid-plane of the plate in the unit cell is broken. Thus, the symmetric and antisymmetric modes hybridize near the original double Dirac cone that introduces the spin-orbit coupling interaction. And their in-phase and out-of-phase hybridization can be used as the effective pseudospin up and down states [26].

Specially, when $\Delta h_S = 3\mu m$ (type-III), the band structure is displayed in Fig. 3(b). The double Dirac cones are split into two single ones and a nontrivial band gap reopens in between them. Compared to the one in Fig. 3(a), it can be seen that the lower Dirac cone combines the lower symmetric and antisymmetric modes and the upper one unites the upper symmetric and antisymmetric modes. The eigenmodes at both Dirac cones are shown in the right panel that illustrate their pairwise hybridization. The spin Chern number can be derived as $C_S^{\uparrow\downarrow}(K/K') = \mp 1/2$ by integrating the Berry curvature of four constituent branches around the valley K (Fig. S12 [35]). Therefore, once assembled with another PPnC with $\Delta h_S = -3\mu m$ (type-IV), the pseudospin-protected edge states would occur at the domain wall as shown in Fig. 3(c). Four spin-locked edge states, namely a forward spin-down ($\downarrow$) pair (the cyan dotted lines) and a backward spin-up ($\uparrow$) pair (the magenta dotted lines), occur in the reopened band gap that is guaranteed by the spin Chern number difference $\pm 1$. Figure. 3(d) shows the propagation of the spin-locked edge states at 1.782MHz localized at the valley K. Both the left-going and right-going waves are excited which is consistent with two available edge states. The refracted patterns at the left and right zigzag terminations predicted by EFCs analysis are depicted in Fig. 3(d). Due to the extreme small wave vectors of $S_0$ and $SH_0$ Lamb modes in the plate, the edge state can only refract into $A_0$ Lamb wave. The wave vectors of the refracted beams (the blue bold arrows) are in good agreement with the out-of-plane displacement field at both terminations.

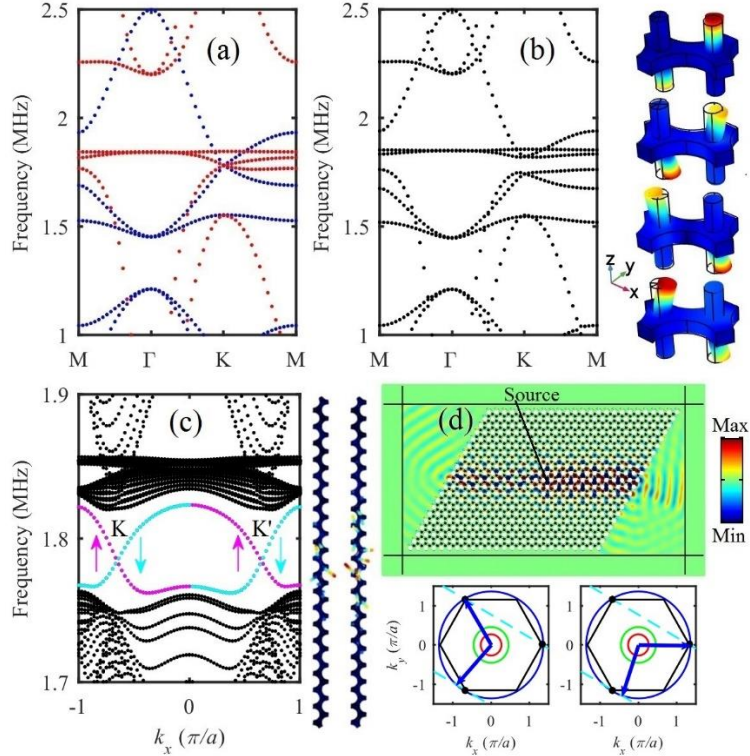

FIG.3: (a) Occurrence of the double Dirac cone. (b) Band structure of the perturbed PPnC and the eigenmodes at both Dirac cones in the descending order. (c) Dispersion curves of the two-layer ribbon supercell constructed by type-III and type-IV PPnCs. (e) Plot of the out-of-plane displacement on the top surface of the plate and the refracted patterns at the left and right zigzag terminations predicted by EFCs analysis.

As we have demonstrated at the beginning, we can involve the perturbation in the diameter to emulate QVHE. Therefore, it provides the opportunity to investigate the pseudospin-valley coupled edge state that utilizes both the pseudospin and valley degrees of freedom. Figure. 4(a) shows the dispersion curves of a two-layer ribbon supercell constructed by arranging PPnCs with $\Delta h_S = -3\mu m$ (type-IV) and $\Delta d_V = -4\mu m$ (type-V) vertically (see the inset). At the valley K, the valley Chern number of type-V PPnC is $C_V^{\uparrow\downarrow}(K) = -1/2$ and the spin Chern number of type-IV PPnC is $C_S^{\uparrow\downarrow}(K) = \pm1/2$, therefore, the magenta dotted line denotes the backward spin-up edge state at DW2 according to the Chern number difference $(-0.5) - 0.5 = -1$. Equally, the cyan dotted line represents the forward spin-down edge state at the valley K'. The purple dotted lines represent the localized modes at the end. Figure. 4(b) displays the dispersion curves of another two-layer ribbon supercell built by arranging PPnCs with $\Delta h_S = 3\mu m$ (type-III) and $\Delta d_V = -4\mu m$ (type-V) horizontally (see the inset). At DW3, a forward spin-up edge state occurs at the valley K' (the magenta dotted line) and a backward spin-down edge state occur at the valley K (the cyan dotted line). Clearly, the propagation of the spin-down edge state at DW2 and DW3 depends on the specific valley. Considering a Y-junction wave guide displayed in Fig. 4(c), the spin-down edge state localized at both valleys [$\psi^{\downarrow}(K/K')$] can propagate at DW1 as discussed in Fig. 3(c), whereas locked to the valley K' [$\varphi^{\downarrow}(K')$] at DW2 and the valley K [$\varphi^{\downarrow}(K)$] at DW3. Therefore, the valley based splitting of the right-going pseudospin-protected (spin-down) edge state can be expected at the junction. In Fig. 4(d), the pseudospin-protected edge states at 1.782MHz at the valley K is selectively excited in the middle of DW1 [37]. The left-going spin-up edge state and the right-going spin-down edge state are both generated. When encountering the junction, the spin-down state turns downwards and propagates along DW3. Vice versa, if the wave vector is locked to the valley K' as shown in Fig. 4(e), the right-going spin-down state only propagates along DW2. Unambiguously, it reveals the valley based splitting of the spin-down edge states.

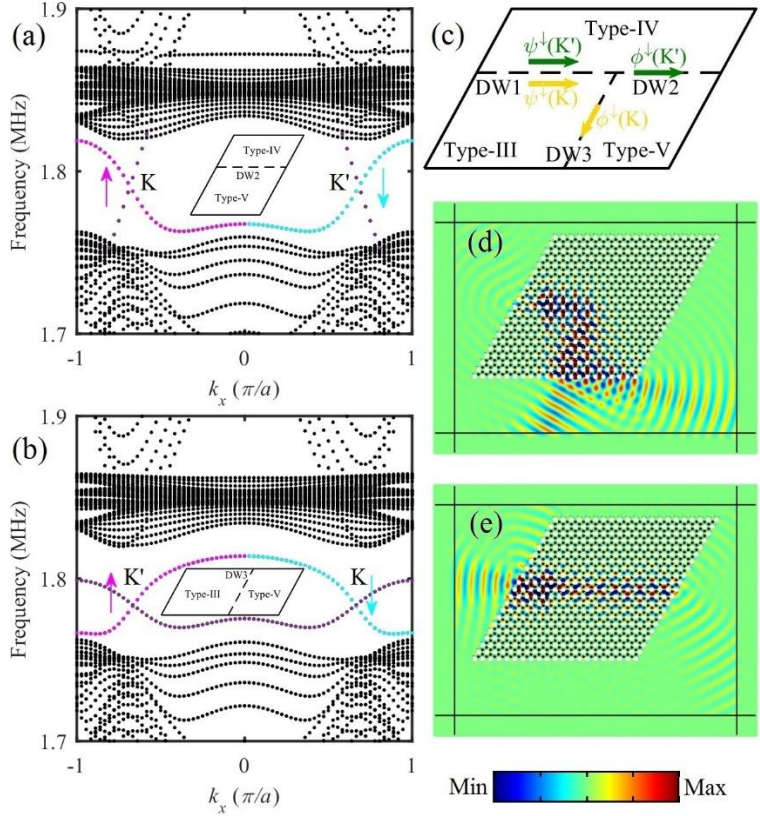

FIG. 4: Dispersion curves of two two-layer ribbon supercells built by (a) type-IV and type-V PPnCs and (b) type-III and type-V PPnCs respectively. (c) Schematic of the Y-junction wave guide and the valley based splitting of the spin-down edge states. Plots of the out-of-plane displacement on the top surface of the plate when the pseudospin-protected edge states are locked to the valleys (d) K and (e) K' respectively.

In conclusion, we demonstrate that the proposed symmetric double-sided PPnC can emulate both QVHE and QSHE by solely imposing different geometric perturbations. The Dirac cones can be judiciously tuned by the parameters of the attached pillars. When mimicking QVHE, we show that both the refractive Lamb modes and angles of the valley-polarized edge state at the zigzag terminations can be tailored by the height of the pillar. Specifically, in the deep subwavelength scale, the edge state turns to be evanescent and can be enhanced by the locally resonant mode at the interface. By optimizing the parameters, a double Dirac cone can be obtained. We evidence the occurrence of the pseudospin-protected edge state in analogy to QSHE. Further, we illustrate the valley-dependent feature of the pseudospin-valley coupled edge state and demonstrate the valley based splitting of the pseudospin-protected edge states in a Y-junction wave guide.